\documentclass[conference]{IEEEtran}

\IEEEoverridecommandlockouts
\usepackage{cite}
\usepackage{amsmath,amssymb,amsfonts}
\usepackage{algorithmic}
\usepackage{graphicx}
\usepackage{multirow}
\usepackage{tabularx}
\usepackage{textcomp}
\usepackage{xcolor}
\usepackage[hidelinks]{hyperref}
\usepackage{xurl}

\begin{document}

\title{Comparative Analysis of SRAM PUF Temperature Susceptibility on Embedded Systems
\thanks{The research reported in this paper has been funded by the Federal Ministry for Climate Action, Environment, Energy, Mobility, Innovation and Technology (BMK), the Federal Ministry for Labour and Economy (BMAW), and the State of Upper Austria in the frame of the COMET Module Dependable Production Environments with Software Security (DEPS) (FFG grant no. 888338) and the SCCH competence center INTEGRATE (FFG grant no. 892418) within the COMET - Competence Centers for Excellent Technologies Programme managed by Austrian Research Promotion Agency FFG.

\copyright~2023 IEEE.  Personal use of this material is permitted.  Permission from IEEE must be obtained for all other uses, in any current or future media, including reprinting/republishing this material for advertising or promotional purposes, creating new collective works, for resale or redistribution to servers or lists, or reuse of any copyrighted component of this work in other works.}
}

\author{
    \IEEEauthorblockN{Martina Zeinzinger, Josef Langer, Florian Eibensteiner,\\Phillip Petz, Lucas Drack}
    \IEEEauthorblockA{\textit{Embedded Systems Lab, University of}\\
    \textit{Applied Sciences Upper Austria}\\
    Hagenberg, Austria\\
    \{martina.zeinzinger, josef.langer, florian.eibensteiner,\\phillip.petz, lucas.drack\}@fh-hagenberg.at}

    \and

    \IEEEauthorblockN{\\Daniel Dorfmeister, Rudolf Ramler}
    \IEEEauthorblockA{\textit{Software Competence Center Hagenberg}\\
    Hagenberg, Austria\\
    \{daniel.dorfmeister, rudolf.ramler\}@scch.at}
}

\maketitle

\begin{abstract}
An SRAM Physical Unclonable Function (PUF) can distinguish SRAM modules by analyzing the inherent randomness of their start-up behavior. However, the effectiveness of this technique varies depending on the design and fabrication of the SRAM module. This study compares two similar microcontrollers, both equipped with on-chip SRAM, to determine which device produces a better SRAM PUF. Both microcontrollers are programmed with an identical SRAM PUF authentication routine and tested under varying ambient temperatures (ranging from 10\,°C to 50\,°C) to evaluate the impact of temperature on SRAM PUF performance. One embedded SRAM works significantly better than the other, even though the two models are closely related. The presented results can be used early in the design process to compare arbitrary on-chip SRAM models and see which is best suited for implementing an SRAM PUF.

\end{abstract}


\section{Introduction}
Physical Unclonable Functions (PUFs) are means to generate a unique identifier for electronic devices. Due to their resemblance to human biometrics, PUFs are sometimes referred to as digital ``fingerprints'' of silicon chips~\cite{herderPufTutorial,gao2020physical,pufTaxonomyMcGrath}. Indeed, similar to a human fingerprint, they are used to differentiate between large numbers of identical devices based on minuscule physical characteristics. These unique characteristics result from manufacturing variations, manifested as physical microstructures or parameters that differ from device to device and cannot be modified or replicated. For example, in SRAM memory, these variations affect the transistors in each memory cell. These transistors have minimal threshold voltage differences, which in turn determine the initial state of the cell after power up. To date, over 40 different approaches to realize PUFs have been proposed~\cite{pufTaxonomyMcGrath}.

In many real-world scenarios, adding additional hardware components is impractical. Thus, implementing PUFs using hardware components that are readily available, such as DRAM and SRAM chips, is beneficial. In environments where embedded systems are prevalent not even these hardware components might be available, requiring a focus on SRAM embedded in the microcontroller. Especially in an industrial context, device authentication for a secure boot process~\cite{VanHerrewege2013,Schaller2014} is of interest, which is a possible use case of PUFs based on embedded SRAM.

Previous works compare different SRAM modules in terms of suitability for SRAM PUFs~\cite{sramAnalysisSchrijenVanDerLeest}. This work presents a comparative analysis of SRAM PUFs implemented using the embedded SRAM of two similar types of STM32 microcontrollers. In particular, the STM32 chips F401RE and F446RE are used. Both chips originate from the same family of microcontrollers, namely the F4 family. They offer a wide range of MCU options and have only small differences in functionality and specifications. Both microcontrollers are widely used in a broad range of embedded system applications. We are interested to see if, despite all the similarities between the two microcontrollers, there is a difference in their suitability for implementing SRAM PUFs.

In contrast to dedicated SRAM modules, it is not always easy to gather information about the embedded SRAM inside microcontrollers. This paper serves to show that even though two microcontrollers might appear to have similar characteristics overall, they can differ in terms of suitability for SRAM PUFs. We describe an automated testing environment in detail, so the presented experiments can be carried out on any standard microcontroller. The methodology shown in this paper can be used early in the design process to compare the suitability of different embedded SRAMs.

This paper discusses SRAM PUF implementations on both the F401RE and the F446RE chip. Temperature tests show the performance of both chips side by side, demonstrating the SRAM PUF's general susceptibility to temperature changes and how both microcontrollers react differently to them. Furthermore, testing both systems with a fuzzy extractor shows the feasibility of developing a complete SRAM PUF authentication system on these microcontrollers. In particular, the experiments show that the choice of chip, even though both options appear similar in functionality, can significantly impact the amount of error tolerance needed in a fuzzy extractor.

\section{Background}

Before going into details of PUFs based on embedded SRAM, in this section, we want to give an overview of physical unclonable functions---focusing on SRAM implementations---, and related noise and error correction mechanisms.

\subsection{Physical Unclonable Functions}

Physical Unclonable Functions (PUFs) are a hardware-based security primitive~\cite{gassend2002silicon,herderPufTutorial,pufTaxonomyMcGrath}.
Minor variations in the manufacturing process of a hardware component cause unintended physical characteristics.
Thus, this unique \emph{digital fingerprint} cannot be cloned easily.
A PUF uses the unique hardware characteristics to provide hardware-specific responses to user-defined challenges.

PUFs can be based on, e.g., Dynamic Random Access Memory (DRAM), which is in wide use as main memory.
DRAM stores bits of information in DRAM cells, which must be refreshed regularly to not lose its charge.
DRAM retention PUFs~\cite{Keller2014,Sutar2016,Xiong2016} utilize this behavior, as they are based on whether individual DRAM cells loose their charge within a specific period of time when they are not refreshed, which varies from cell to cell.
Another possible implementation is the DRAM Latency PUF~\cite{Kim2018}, which deliberately violates timings for certain DRAM operations, e.g., reading from DRAM, to cause errors that are used for the PUF response.

\subsection{SRAM PUFs}
One specific way of implementing a PUF is the so-called SRAM PUF. It is based on the start-up behavior of SRAM chips, in particular the preferred start-up values of the individual SRAM cells. Essentially, powering any given SRAM chip causes a random pattern to appear among its individual cells, where some cells power up to a 1 and the others to a 0~\cite{sramPufFingerprintHolcomb, sramPufGuajardo, herderPufTutorial, pufTaxonomyMcGrath, sramRFIDHolcomb}. Except for a small amount of noise, the emerging pattern is the same each time that the chip is powered on. Each SRAM chip shows a different pattern, making individual chips distinct, and allowing for unique authentication.

To show the noise that emerges when powering the same chip multiple times, the start-up pattern of an SRAM chip is illustrated in \autoref{fig:sramFingerprint}. Specifically, the figure shows the average probability over a set of 100 readings of powering up to 1 for each individual cell. If this probability is 0\,\% or 100\,\%, the cell is strong, showing the same value after each power-up. If the probability is somewhere in between, the cell is weak and less useful for the identification of the chip. Rounding all the resulting probability values to 0 or 1 gives the most likely start-up pattern that can be obtained and thus the known fingerprint of the SRAM~\cite{sramPufFingerprintHolcomb}. Given an SRAM chip, taking a random fingerprint $F_a$ and comparing it to its known fingerprint $F_K$ results in less noise than comparing it to any other random fingerprint $F_b$.

\subsection{Noise and Error Correction}
\label{sec:noiseAndErrorCorrection}
\begin{figure}
    \makebox[\linewidth][l]{%
        \includegraphics[width=1\linewidth]{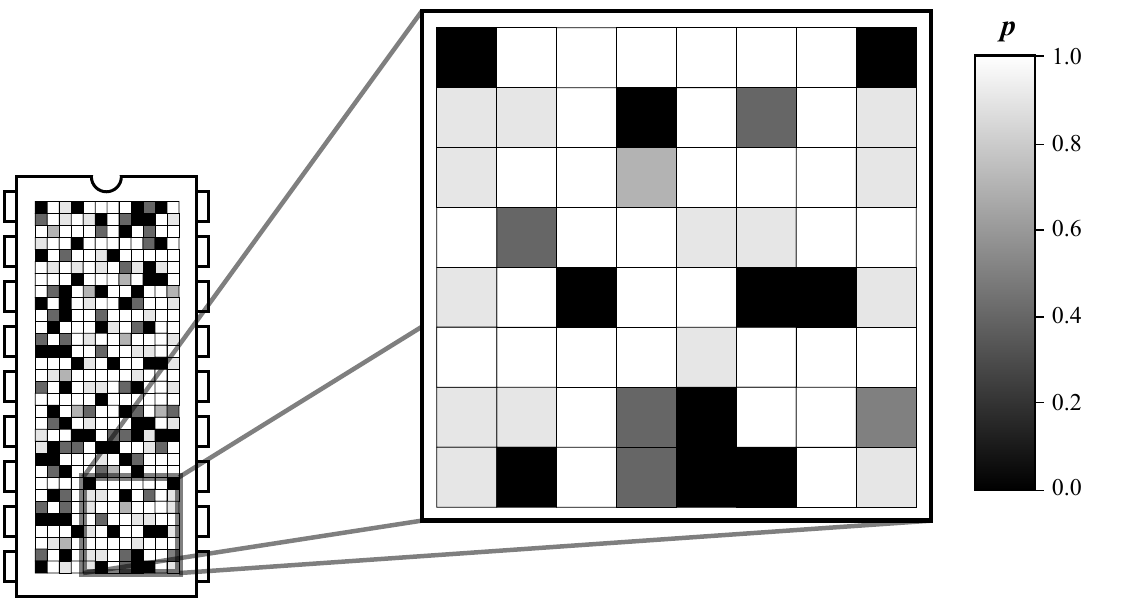}%
    }%
    \caption{The start-up pattern of an SRAM chip, averaged over a 100 readings. The shading of each cell represents the cell's probability of powering up to 1. White and black cells have a strong probability of 100\,\% and 0\,\%, respectively. Gray cells can be considered unreliable, making them so-called weak cells.}
    \label{fig:sramFingerprint}
\end{figure}

As with traditional biometrics, any pattern generated from an SRAM PUF includes some degree of noise. This stems from the fact that a minority of SRAM cells behave randomly, represented in grey in \autoref{fig:sramFingerprint}. In other words, no two readings are exactly the same. Correcting this noisy input is necessary in order to use the fingerprint in a cryptographic authentication system. Here, fuzzy extractors~\cite{fuzzyDodis} based on error-correcting codes (ECC) have become established as the preferred software-based approach of dealing with noisy input~\cite{herderPufTutorial, sramAnalysisSchrijenVanDerLeest, pufsAtAGlanceHolcomb, fuzzyKangEfficient, fuzzyImplementationKang}.

Fuzzy extractors use a two-stage process, where a reference fingerprint is first enrolled with the fuzzy extractor, producing helper data. A new fingerprint can later be authenticated against the reference, using said helper data. This authentication allows for a margin of error to exist in the fingerprint, thus accepting samples that differ from the reference. Most notably, the reference fingerprint is destroyed after enrollment and must therefore not be stored in non-volatile memory (NVM), which is unsafe. Helper data, on the other hand, can be stored safely in NVM as they must not leak any information about the original reference.

Taken together, the typical authentication procedure in the field involves the device going through a regular power-up. During power-up, SRAM forms its characteristic fingerprint in the uninitialized memory. Early in the boot process, firmware reads this fingerprint and passes it to the fuzzy extractor. If the fingerprint is reasonably close to the original, a PUF-based identifier is created and sent to an authentication authority, remote or on-site.

\section{SRAM PUFs in Embedded Memory}
\label{sec:srampufOnEmbeddedMemory}
In principle, any SRAM chip is able to host an SRAM PUF. This paper specifically covers embedded SRAM chips, such as those found in most microcontrollers. Here, some specific aspects should be considered when designing an authentication system.

\subsection{Start-Up Behavior}
\label{sec:startupBehavior}
SRAM PUFs produce their unique response only when they are turned on, i.e., when a voltage is applied to the previously unpowered chip. This constrains the area of application for SRAM PUFs when they are implemented on embedded SRAM, since the chip cannot be turned off and on at will. Performing a reset for the sake of authentication would be counterproductive. In contrast to other PUF types, the SRAM PUF can only produce a response when the SRAM is fully disconnected from voltage, so that its cells can lose their charge, and subsequently be repowered.

Importantly, SRAM PUFs can be implemented with memory that is later re-used by the user program, given some precautions. Whether the memory area is re-used by the user program or not, the PUF response must be overwritten after authentication. Otherwise, one could simply read the PUF response from memory, compromising the unique fingerprint. Overwriting the PUF response also serves to accomplish anti-aging, a technique used to improve long-term stability of the SRAM PUF, as presented by Maes and van der Leest~\cite{sramPufAgingMaes}.

\subsection{Embedded SRAM Reset Behavior}
\label{sec:resetBehavior}
While it is typically recommended to dedicate a separate memory area to the SRAM PUF only, this area can also be re-used by the user program if the need arises. In this case, however, the aging of the SRAM and the reset behavior of the device must be taken into account. Generally, while performing a system reset, ARM microcontrollers do not cut off the voltage powering the embedded SRAM~\cite{stm32f401reRefMan, stm32f446reRefMan}. The respective devices might perform a reboot of the microcontroller, but the values stored in the SRAM are not erased. Consequently, the PUF would not be able to recover its response and fail the authentication. It is therefore crucial that the device is fully reset in any possible situation, e.g., by physically disconnecting the battery upon reset.

In the same vein, voltage dips must be considered. A dip in voltage could trigger a system reset, e.g., through brown-out detection. This is another case when the system might enter the undesired state where the microcontroller resets while the SRAM keeps its old values.

\subsection{Sensitivity to Environmental Factors}
\label{sec:temperature}
Along with the effects of supply voltage ramp-up time~\cite{sramPufRampUpTime} and so-called aging caused by negative bias temperature instability~\cite{sramPufAgingMaes}, temperature changes can have a substantial impact on the performance of SRAM PUFs~\cite{sramPufFingerprintHolcomb, sramAnalysisSchrijenVanDerLeest, sramPufIntrinsicID, sramRFIDHolcomb}.

Most notably, changes in temperature cause individual cells to change their behavior. Some cells might be stable at room temperature while becoming unreliable when temperatures rise or fall. What is most important is that the absolute temperature of operation is irrelevant to the authentication of the device. What matters is solely the temperature difference between the time of enrollment and the time of authentication. Since enrollment happens in the factory where the device is programmed, the fingerprint emerging at the current temperature at that time is the baseline of comparison. In the following experimental results, we discuss both the absolute noise present at the different temperature points as well as the relative noise to the reference temperature.

\section{Experiments}
\label{sec:experiments}
We designed the following experiments to be as reproducible as possible. They can be reproduced with any embedded microcontroller capable of hosting an SRAM PUF. The experimental data presented here serves to show the temperature susceptibility of SRAM PUFs by comparing two different, yet very similar, models of embedded microcontrollers.

\subsection{Used Microcontrollers}
\label{sec:microcontrollers}
\begin{figure}
    \centering
    \includegraphics[width=\linewidth]{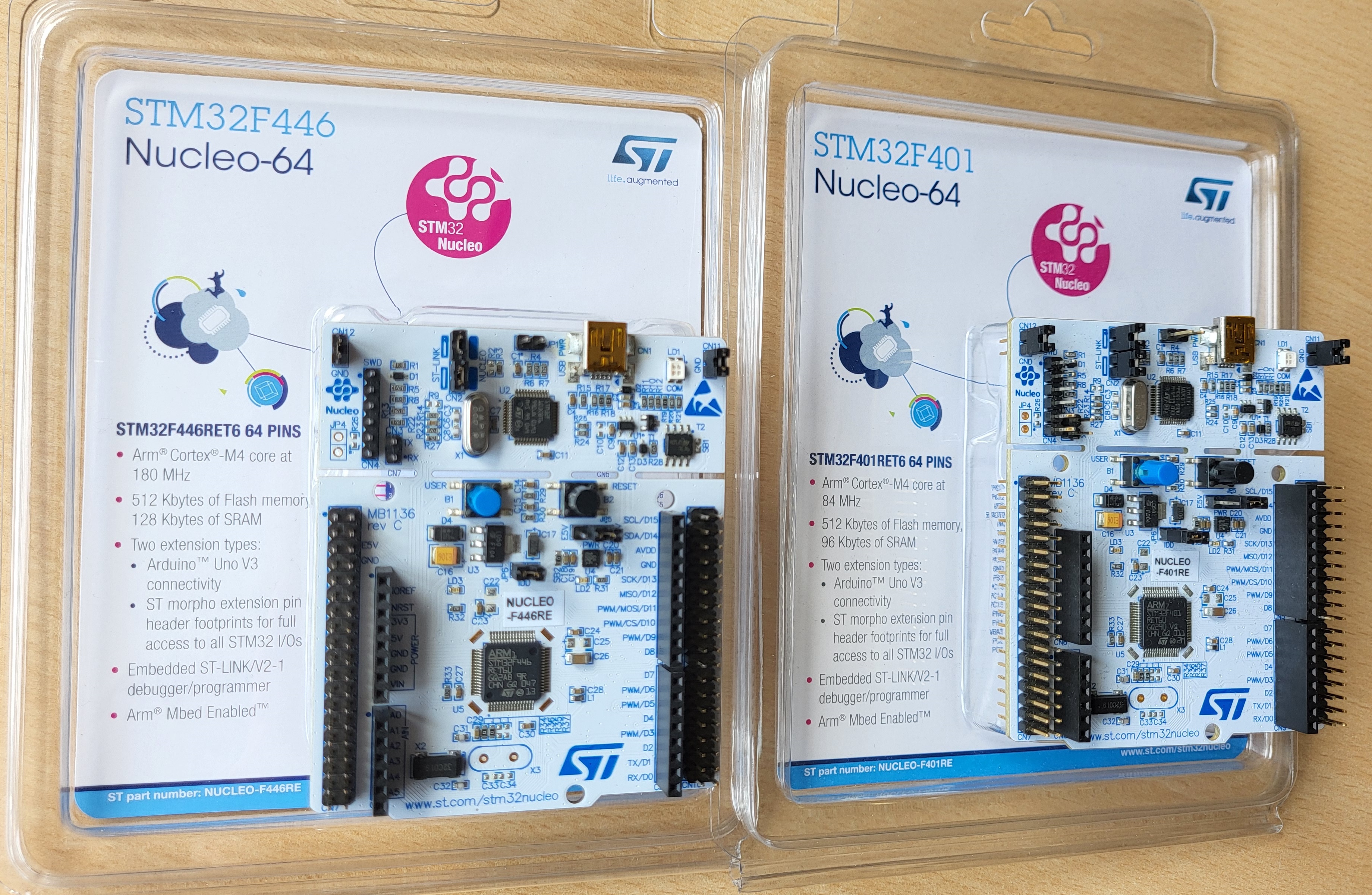}
    \caption{The two tested versions of STM Nucleo boards: STM32F446RE on the left side and STM32F401RE on the right side.}
    \label{fig:uCBoards}
\end{figure}

We chose two related board types from the STM Nucleo development board range for conducting the experiment: the STM32F401RE boards with 96\,KiB embedded SRAM~\cite{stm32f401reRefMan} and the STM32F446RE boards with 128\,KiB embedded SRAM~\cite{stm32f446reRefMan} (see \autoref{fig:uCBoards}), both fabricated using a 90\,nm process~\cite{stm32F401page, stm32F446page}. Further features of each board are summarized in \autoref{tab:uCBoards}. The selection comprised of 14 devices of each type. Both device types belong to the same microprocessor family, equipped with ARM Cortex-M4 cores. To put both devices into perspective, the F401RE is an entry level device running at 84\,MHz core speed, while the F446RE is positioned as a high-performance option operating at 180\,MHz.

\subsection{Experimental Setup}
\label{sec:experimentalSetup}
The temperature experiments were conducted in a compact temperature chamber, capable of reaching temperatures from 10\,°C to 50\,°C. This chamber is custom-built for small device testing and is able to hold temperatures stable, even when the contained devices are under load. 

For each test run, we brought the devices inside the chamber to the desired temperature and kept them there for several minutes before we recorded the experiment data, i.e., the SRAM PUF responses. We did not mix the two device types during test runs. This means that for each run, we populated the chamber with 14 boards of the same type. For recording of the SRAM PUF responses, we used a single development PC to which all 14 boards were simultaneously connected via USB hubs. We simultaneously switched on and off the boards by physically interrupting the USB connection.

\begin{figure}
    \centering
    \includegraphics[width=\linewidth]{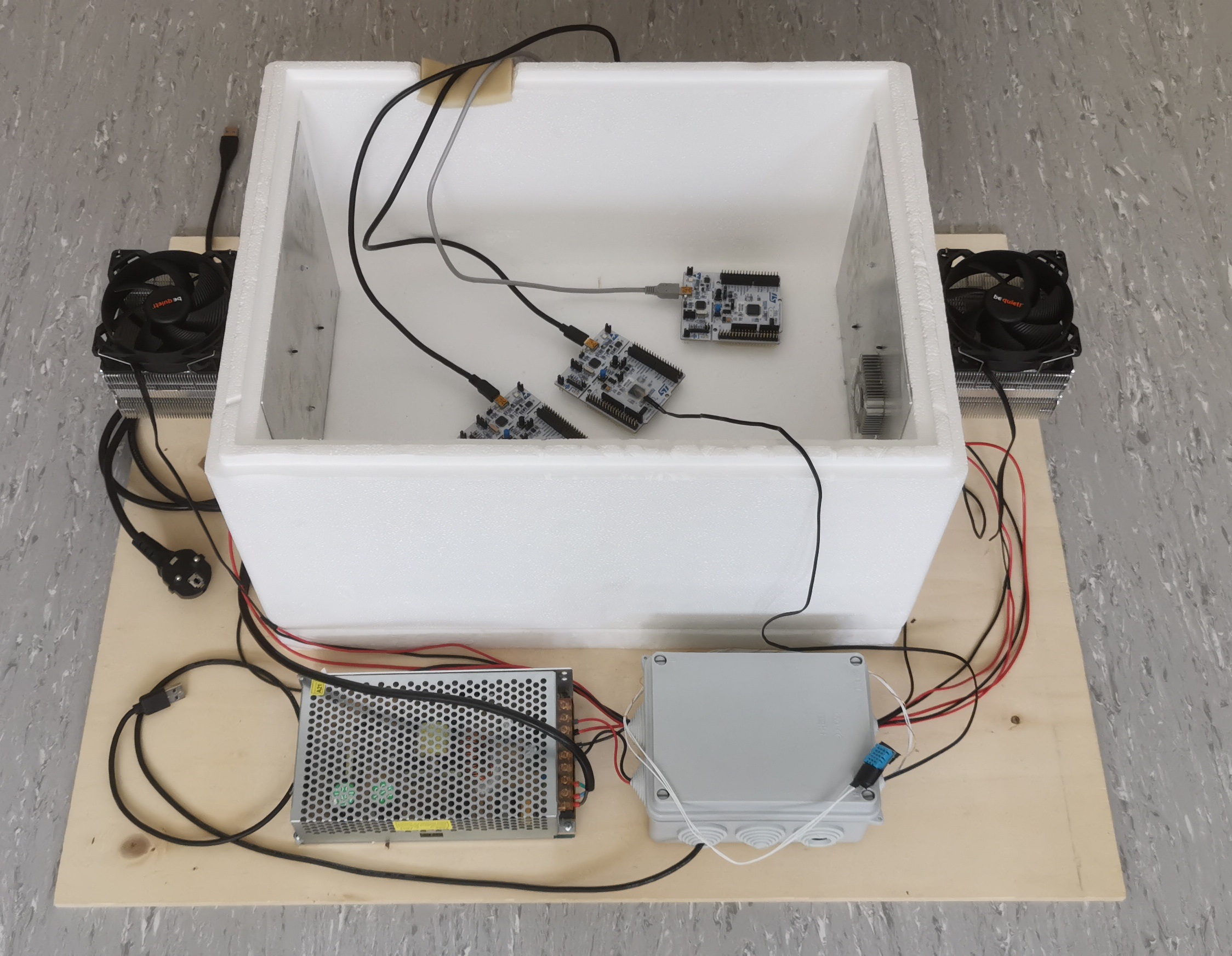}
    \caption{Low-cost climate chamber for tests in a temperature-stable environment. The styrofoam chamber can accommodate devices up to the size of ATX mainboards. Heat is transferred from the inner fan and the inner aluminum plate to the outer CPU fan by two Peltier elements on each side. Depending on the waste heat from the electronic devices inside, precise control of the target temperature from 0\,°C to 55\,°C is possible.}
    \label{fig:climateChamber}
\end{figure}

\begin{table}[b]
    \caption{Main features of the two evaluated Nucleo-64 boards by STMicroelectronics.}
    \begin{center}
        \begin{tabularx}{\linewidth}{@{}llccc@{}}
            \hline
            \rule{0pt}{3ex} & & & \textbf{STM32F401RET6} & \textbf{STM32F446RET6} \\ \hline
            \rule{0pt}{3ex} & Processor & & Arm\textsuperscript{\textregistered} Cortex\textsuperscript{\textregistered}-M4 core & Arm\textsuperscript{\textregistered} Cortex\textsuperscript{\textregistered}-M4 core \\
            \rule{0pt}{3ex} & Clock Speed & & 84\,MHz & 180\,MHz \\
            \rule{0pt}{3ex} & Flash & & 512\,KiB & 512\,KiB \\
            \rule{0pt}{3ex} & RAM & & 96\,KiB SRAM & 128\,KiB SRAM \\ 
            \rule{0pt}{3ex} & Product Line & & Access Line & Foundation Line \\
            \rule{0pt}{3ex} & \multirow{2}{6em}{Temperature Range} & & \multirow{2}{*}{\centering{-40\,°C to +105\,°C}} & \multirow{2}{*}{\centering{-40\,°C to +105\,°C}} \\ \\
            \\ \hline
        \end{tabularx}
        \label{tab:uCBoards}
    \end{center}
\end{table}

\subsection{Metrics and Notation}
\label{sec:metrics}
To quantify SRAM PUF performance, a variety of metrics exist in the literature. Among the most important of them is the fractional Hamming distance (\textsl{FHD}). Given two fingerprints of the same size, \textsl{FHD} describes how many bits differ between them as a percentage. Generally, it is used as an indication of how much average noise an SRAM PUF produces. While noise stays well under 10\,\% for most SRAM modules, higher values are certainly plausible when temperature conditions vary~\cite{sramAnalysisSchrijenVanDerLeest}. This is the noise that the error correction methods we mentioned in \autoref{sec:noiseAndErrorCorrection} have to correct.

Among other metrics, the fractional Hamming distance further serves to calculate the reliability and uniqueness of SRAM PUFs~\cite{pufPerformanceMeasuresMaiti}. Reliability indicates how reliably a single chip reproduces the exact same response with each power-up. To calculate the reliability of one SRAM model, many instances of that SRAM must be tested multiple times. Therefore, reliability is also referred to as intra-class Hamming distance, or $\textsl{HD}_{\textit{intra}}$~\cite{herderPufTutorial, pufAuthenticationDeutschmann}.

$\textsl{FHD}$ is also needed to calculate SRAM PUF uniqueness, which quantifies how much two fingerprints differ from each other~\cite{pufPerformanceMeasuresMaiti}. A uniqueness value of 0\,\% would mean that a fingerprint was compared to an exact copy of itself. When comparing SRAM instances, this metric ideally evaluates to 50\,\%, since this is what would be expected when comparing two truly random strings~\cite{pufPerformanceMeasuresMaiti}. Uniqueness is also referred to as inter-class Hamming distance, or $\textsl{HD}_{\textit{inter}}$~\cite{herderPufTutorial, pufAuthenticationDeutschmann}.

Reliability and uniqueness are the key metrics of comparison for the following experimental results. For a rigorous definition and classification of the used PUF performance metrics, refer to Maiti et al.~\cite{pufPerformanceMeasuresMaiti}.

\section{Results and Discussion}

In this section, we present the results of the experiments described in \autoref{sec:experiments} and discuss their implications.

\subsection{Measurement Series and Temperatures}
\label{sec:temperatures}
\begin{figure*}
    \centering
    \includegraphics[width=\textwidth]{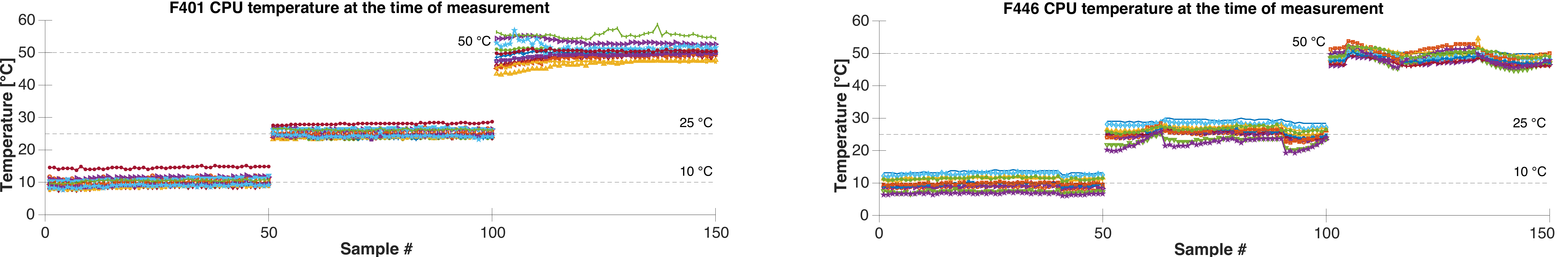}
    \caption{Temperature for each collected sample as measured by the internal temperature sensor of each board. Each line and color stands for a particular board, resulting in 14 lines in total for each graph. The left graph shows the values for the F401RE, while the right graph shows values for the F446RE. As indicated by the x-axis, 50 samples were taken at each temperature point.}
    \label{fig:temperatures}
\end{figure*}

We carried out multiple test runs with the boards inside the climate chamber (see \autoref{fig:climateChamber}). In total, we conducted 150 measurement runs for each board type. \autoref{fig:temperatures} shows the temperatures measured by the internal temperature sensors of the boards for each individual reading. Evidently, the internal temperature sensors do not provide an accurate assessment of the actual temperatures as they vary substantially from one reading to the next. However, what matters in this context is not the actual temperature that the measurements were taken at but the difference between the three temperature settings. The graphs in \autoref{fig:temperatures} show that the measured temperatures match the goal temperatures of 10\,°C, 25\,°C and 50\,°C adequately.

\begin{table}[b]
    \caption{Average noise across temperatures. See \autoref{fig:intraHD25} for the associated diagrams.}
    \begin{center}
        \begin{tabular}{@{}lcccc@{}}
            \hline
                   & & $\textsl{FHD}_{\textit{avg10}}$ & $\textsl{FHD}_{\textit{avg25}}$ & $\textsl{FHD}_{\textit{avg50}}$ \\
                   & & [\%]   & [\%]     & [\%] \\ \hline
            F401RE & & 5.29   & 3.87     & 5.35 \\ 
            F446RE & & 6.79   & 4.24     & 7.72 \\ \hline
        \end{tabular}
        \label{tab:intraHD25}
    \end{center}
\end{table}

\begin{figure*}
    \centering
    \includegraphics[width=\textwidth]{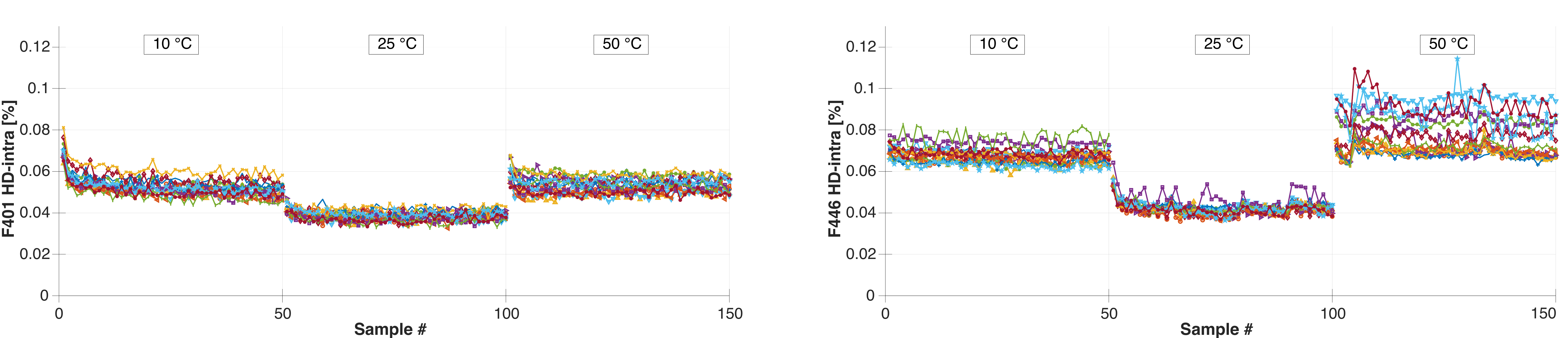}
    \caption{$\textsl{HD}_{\textit{intra}}$ of both SRAM types in comparison. Each data point stands for the \textsl{FHD} between the corresponding fingerprint and the reference fingerprint we took at 25\,°C. The data points correspond to those in \autoref{fig:temperatures}. Again, there are 14 lines per measurement series. Averaging all 14 lines gives the values found in \autoref{tab:intraHD25}. The left graph shows the results for the F401RE, while the right graph shows results for the F446RE.}
    \label{fig:intraHD25}
\end{figure*}

\subsection{Comparison with a Reference Fingerprint}
\label{sec:resultsIntraHD25}

In an authentication scenario, the SRAM PUF is enrolled at a certain temperature, e.g., room temperature, and authenticated later at a different temperature. It is therefore reasonable to first derive a reference fingerprint for each board at a specified temperature. This is the baseline of comparison for the fingerprints gathered at all other temperatures. This is important as the PUF must be guaranteed to work securely even under adverse conditions, i.e., when temperatures approach the upper or lower bound of what the SRAM is specified for.

The following experiment shows this scenario for temperature points at 10\,°C, 25\,°C and 50\,°C. In all test cases, 25\,°C is the baseline against which we compare all fingerprints. To achieve this, we aggregated a reference fingerprint at 25\,°C by taking 50 measurements and averaging them, resulting in the known fingerprint $F_{K}$. We repeated these measurements for each development board. We ended up with 28 known fingerprints in total, 14 from the F401RE and 14 from the F446RE, one for each device. Then, we read 150 new fingerprints from each device at the temperatures displayed in \autoref{fig:temperatures}. We compared these new fingerprints to the known fingerprints by calculating their \textsl{FHD}. As a final result, \autoref{fig:intraHD25} shows the $\textsl{HD}_{\textit{intra}}$ for each single board, grouped by board type. Averaging the noise values of all the tested chips gives the values listed in \autoref{tab:intraHD25}, separated by board type and temperature.

As we can gather from these results, the two chip types differ substantially from each other in terms of their average noise level. The F446RE exhibits higher levels of noise across the board. With the F446RE, the standard deviation at 50\,°C also seems to be much greater than that of the F401RE, suggesting that its in-class variance increases with temperature. Still at 50\,°C, the F446RE's highest recorded $\textsl{HD}_{\textit{intra}}$ is 11.4\,\%, while noise levels on the F401RE rarely exceed 6.5\,\%. The F401RE performs more consistently overall.

Comparing the values from \autoref{tab:intraHD25}, the noise present at 25\,°C in the F401RE is 8.7\,\% lower than that of the F446RE. At 10\,°C, the values are 22.1\,\% lower, while at 50\,°C, they are 30.7\,\% lower. This is a substantial difference. Additionally, it can be assumed that this pattern will continue as temperatures are further increased or decreased.

Looking at these results, we can conclude that the F401RE is better suited as basis for an SRAM PUF on account of the reduced noise level of around 8.7\,\% at room temperature. This reduction alleviates the demand for error correction. When considering that the difference between noise levels becomes even more severe when temperatures deviate further from the reference temperature, the F401RE is clearly the better choice. Nevertheless, both chips show the same general behavior, namely that noise rises with the relative temperature difference between the device's current temperature and its temperature at enrollment. These results go in line with those from larger studies~\cite{sramPufIntrinsicID, sramAnalysisSchrijenVanDerLeest}. Compared with the results from the mentioned studies, the F401RE performs well. The performance of the F446RE is mediocre, but nonetheless viable.

\subsection{Reliability}
\label{sec:reliability}
In the previous section, we analyzed the PUF behavior when fingerprints taken at high or low temperatures are compared to a reference that was taken at 25\,°C. Besides, what is also of interest is the absolute reliability at said temperature points.

\begin{figure}
    \centering
    \includegraphics[width=\linewidth]{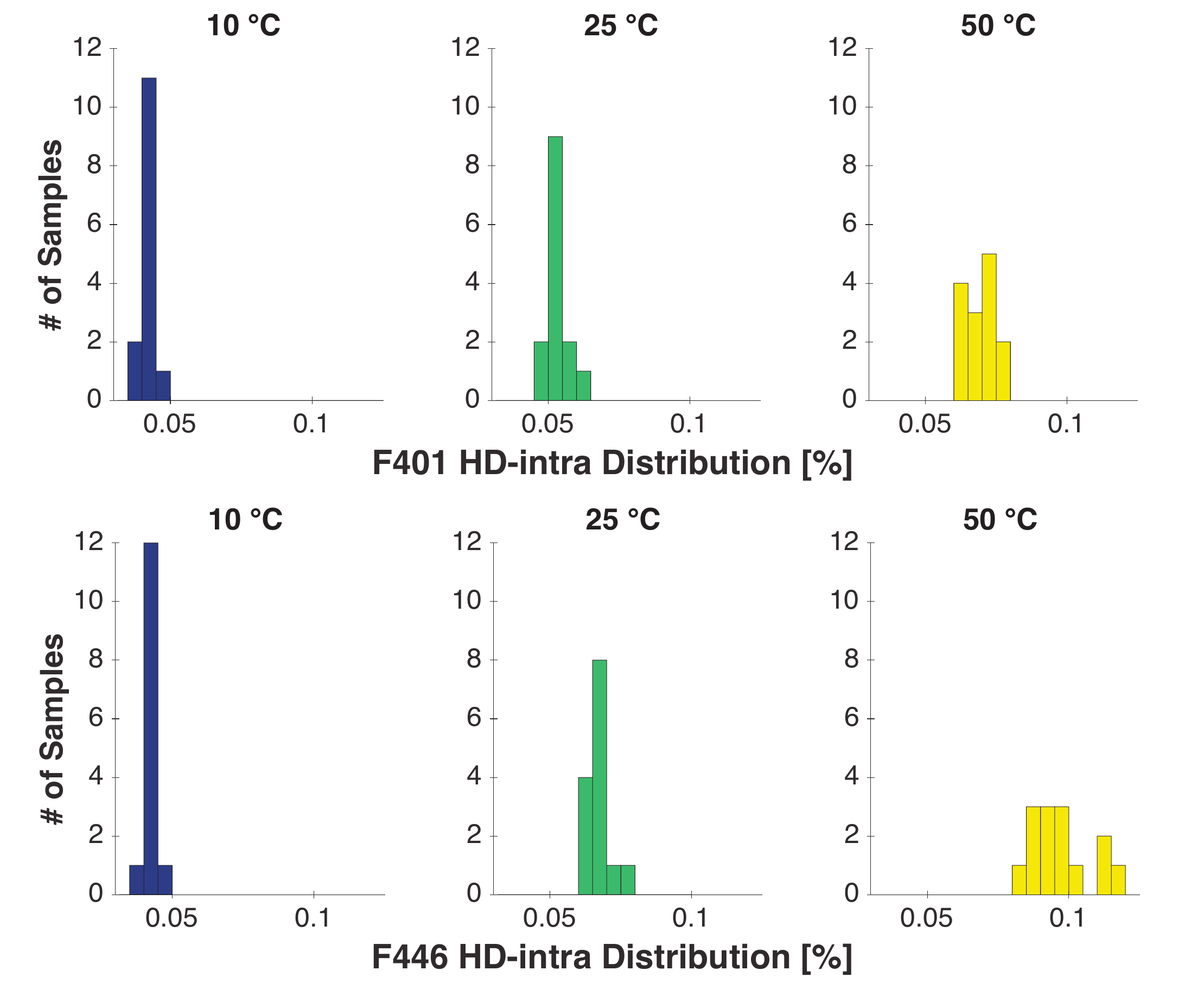}
    \caption{Noise distribution of the two chip types at fixed temperatures. The collection of graphs at the top shows the F401RE, while the F446RE is at the bottom. For each chip type, 14 boards were used. From each board, 50 samples were taken per temperature.}
    \label{fig:intraHD1}
\end{figure}

\begin{figure}
    \centering
    \includegraphics[width=\linewidth]{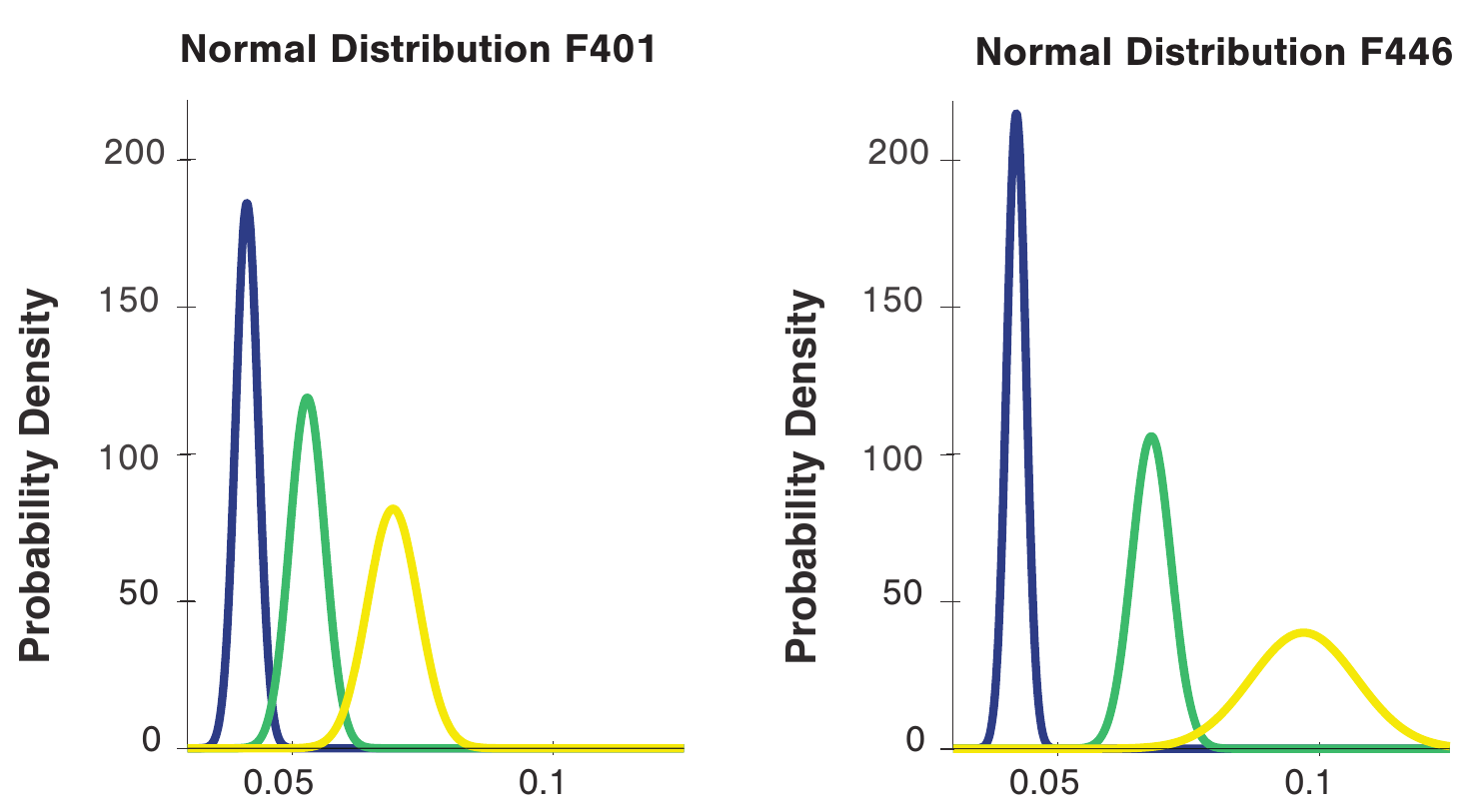}
    \caption{These graphs show the approximated distributions of all three temperature points side by side. Graph a) on the left shows the normal distributions for the tested temperatures on the F401RE, while b) at the right gives this information for the F446RE.}
    \label{fig:intraHD2}
\end{figure}

\autoref{fig:intraHD1} and \autoref{fig:intraHD2} display how temperature affects the reliability metrics. The diagrams for both chip types show a similar behavior, as the reliability of both variants is highest when temperatures are low. When temperatures rise, reliability declines---which means that more bits become unreliable by showing unpredictable behavior.

Comparing the two graphs, it is clearly visible that the F446RE shows higher levels of noise by a substantial margin. It shows almost twice as much noise at 50\,°C compared to the F401RE. Generally, and most notably on the F446RE, standard deviation of noise distribution also increases with temperature.

In contrast, both models seem to behave remarkably stable when chilled to 10\,°C.

\subsection{Uniqueness}
\label{sec:uniqueness}
\begin{figure}
    \centering
    \includegraphics[width=\linewidth]{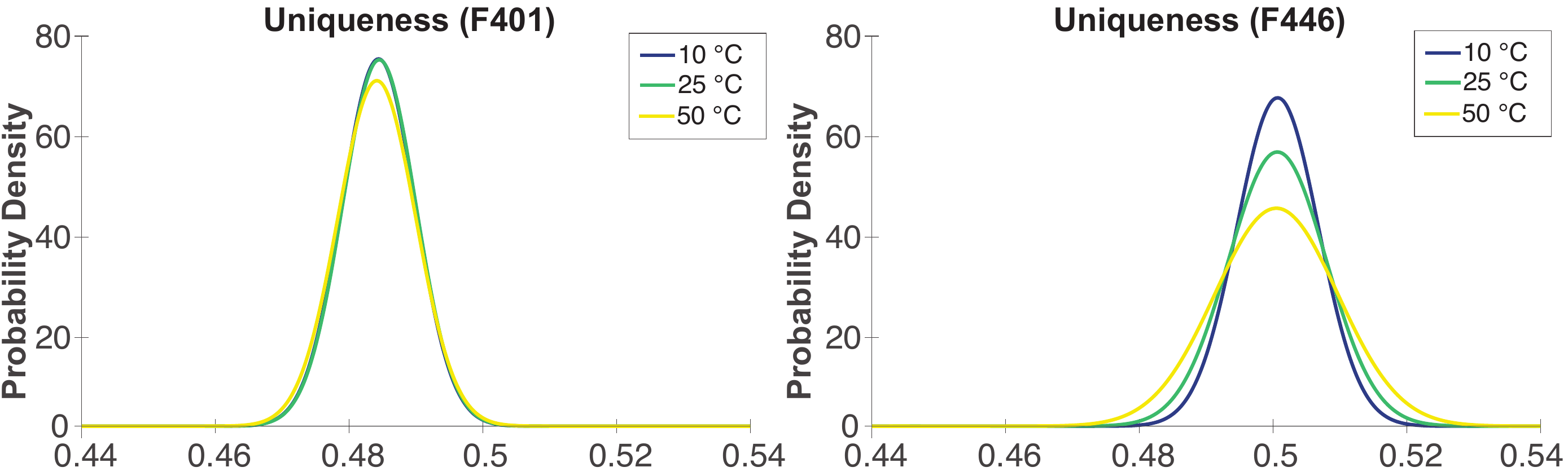}
    \caption{Normal distributions of uniqueness for the F401RE and F446RE side by side. Scores for the F401RE do not vary substantially with temperature, therefore the lines largely overlap.}
    \label{fig:uniquenessf401vsf446}
\end{figure}

Uniqueness describes how different the individual fingerprints are between different specimens of the same SRAM models. \autoref{fig:uniquenessf401vsf446} shows that the F446RE actually scores slightly better in this regard, staying close to the ideal 50\,\%. However, it shows an increase in standard deviation when temperature rises, whereas the F401RE stays largely constant throughout. This increased standard deviation can be attributed to the behavior shown in \autoref{fig:intraHD1} and \autoref{fig:intraHD2}, since the F446RE in general shows elevated levels of noise at 50\,°C.

\subsection{Implications for Error Correction}
\label{sec:errorCorrection}
Error correction for PUFs is usually achieved by the use of fuzzy extractors~\cite{herderPufTutorial, sramAnalysisSchrijenVanDerLeest, sramPufFingerprintHolcomb, fuzzyKangEfficient, fuzzyImplementationKang}. To date, no real alternative to fuzzy extractors has emerged. This begs the question which implementation is the best for embedded systems and their strictly limited resources. Taking into account the presented temperature behavior of the SRAM fingerprints, any fuzzy extractor design must be prepared for the worst case in terms of environmental conditions. They must provide sufficient tolerance in order to be able to correct bit errors in the SRAM PUF response, even when temperatures reach the limits of specification. As the presented experiments show, conclusions about the worst case conditions can only be determined empirically for each individual SRAM model and can differ considerably, even in SRAM architectures that seem to be closely related in design. Temperature tests, such as those described in this paper, are therefore a necessary step in SRAM PUF development and must also include the fuzzy extractor design.

To study the impact of temperature dependence on a fuzzy extractor empirically, we implemented a fuzzy extractor in C based on a construction of Canetti et al.\cite{fuzzyCanetti}. This construction is not based on traditional ECC, but rather on digital lockers~\cite{smartLockerCanetti}, using a sample-then-lock mechanism. It relies on obscuring the secret, in this case the reference fingerprint, with a large number of random nonces. The construction has adjustable parameters with values for the desired Hamming distance, reproduction error and cipher security. Adjusting the parameters allows the fuzzy extractor to be more or less tolerant in terms of bit errors, i.e., allowing a larger Hamming distance between the reference and any new fingerprint. With more tolerance comes a larger number of random nonces needed. So, the trade off of a more tolerant fuzzy extractor is an increase in the size of the helper data.

Deploying a fuzzy extractor on an embedded system implies that its tolerance is kept to the necessary minimum due to the limited memory size. It is of interest to note once again that the size of helper data increases with the required error tolerance. Even though the construction of Canetti et al. does not score at this point, it is sufficiently suitable for our purposes.
Resulting from the performed tests, which involved fingerprints of 16 bytes and a fixed reproduction error of $10^{-3}$, the size of the helper data for a Hamming error of 4 bits amounted to 30\,KiB of data. In order to correct 5 faulty bits, the size of the helper data increases to 81\,KiB. Finally, adjusting the Hamming error to 8 bits results in a substantial increase in needed memory, with 1600\,KiB of helper data. So, with the construction of Canetti et al., increasing the desired Hamming distance by one bit, while keeping the other parameters the same, results in an increase in helper data size of 265\,\%. Real world implementations of SRAM PUFs require fingerprints with a size of hundreds of bytes. Larger fingerprints naturally require a higher error tolerance in terms of Hamming distance, so this exponential growth makes the construction largely impractical. This suggests that the traditional ECC-based methods are better suited to this task because of the amount of memory required in a final product. Clearly, the amount of helper data that this implementation requires is beyond the capabilities of modern embedded systems. However, as already mentioned, for our purposes - the demonstration of the different temperature sensitivity in very similar microcontrollers - the chosen fuzzy exractor is sufficient. Since no memory has to be reserved for an application, we can use it sufficiently well to show our results. 

Therefore, we tested it with data from the temperature experiments, using fixed size fingerprints of 16 bytes. We set the allowed reproduction error to $10^{-3}$ and allowed 5 bits of Hamming error. This corresponds to 3.9\,\% of allowed Hamming error. First, we enrolled the fuzzy extractor with a reference fingerprint aggregated at 25\,°C. Then, we attempted replication with random fingerprints we also took at 25\,°C. The construction could easily reconstruct the fingerprints. However, when attempting the same with fingerprints that were recorded at 10\,°C or 50\,°C, the fuzzy extractor failed to reconstruct the fingerprint. This is due to the elevated noise levels which are demonstrated in \autoref{fig:intraHD25}. Readings from both the F401RE and the F446RE behaved the same.

Using the same reference fingerprint, but with an increased tolerance of 8 bits of Hamming distance, leads to a different behavior. It was now able to reconstruct its reference fingerprint, even when the reproduction material used was recorded at 10\,°C or 50\,°C. Noticeably, processing time increased greatly, which is another drawback of having to store a large amount of helper data.

For these tests, we used fingerprints of 16 bytes. Comparatively, this is a tiny size for SRAM PUF systems, where byte sizes usually lie in the hundreds. The tiny size of 16 bytes was necessary due to constrains in processing power. Using a size of 32 bytes, as an example, would increase the processing time from minutes to hours, which was impractical to test. A different solution is therefore needed in order to bring the construction of Canetti et al. onto embedded systems. For now, traditional ECC-based solutions have the edge.

\section{Conclusion}
\label{sec;conclusion}
We found both chips, F401RE and F446RE, to be viable for implementation of SRAM PUFs using their embedded SRAM. Important remarks about the implementation of SRAM PUFs on the embedded SRAM are presented especially in terms of their start-up and reset behavior. The susceptibility of SRAM PUFs to environmental factors is demonstrated through temperature experiments, showing that PUF performance changes with the ambient temperature. In particular, we carried out temperature tests at 10\,°C, 25\,°C and 50\,°C.

Fundamentally, we confirmed the findings of previous research~\cite{sramPufFingerprintHolcomb, sramPufIntrinsicID, sramAnalysisSchrijenVanDerLeest} by the presented results. Most importantly, the results summarized in \autoref{fig:intraHD25} show the expected behavior when fingerprints from different temperatures are compared with a reference taken at a fixed temperature.

What is most surprising is that the two chips, closely related and stemming from the same product family, performed so differently when utilized for SRAM PUFs. One would expect them to be much closer in terms of average noise, however we found the F401RE to be significantly better in this critical aspect compared to the F446RE. No statement can be made here about the causes for these deviations. A closer investigation of the reasons for the different behavior is reserved for future work.  The same applies to the comparison between specimens of the same microcontroller from different production batches or also to the comparison with other microcontrollers from the same series.

In addition to the tests done on the SRAM chips themselves, we presented and discussed an implementation of a fuzzy extractor based on a construction by Canetti et al.~\cite{fuzzyCanetti}. We found the design to be working, but inadequate for use in embedded systems in terms of needed helper data size and processing time. For now, traditional ECC-based algorithms have the edge over Canetti et al.'s digital lockers~\cite{smartLockerCanetti} in this regard. Given a sufficient reduction of the size of needed helper data, this might change in the future.

\bibliographystyle{IEEEtran}
\bibliography{references.bib}

\end{document}